# A one-dimensional model for water desalination by flow-through electrode capacitive deionization

Eric N. Guyes,[1] Amit N. Shocron,[1] Anastasia Simanovski,[1]
P.M. Biesheuvel,[2] Matthew E. Suss[1,*]

[1]Faculty of Mechanical Engineering, Technion – Israel Institute of Technology, Haifa, Israel.
[2]Wetsus, European Centre of Excellence for Sustainable Water Technology, Leeuwarden, The Netherlands.
[*]*mesuss@tx.technion.ac.il*

## Abstract

Capacitive deionization (CDI) is a fast-emerging water desalination technology in which a small cell voltage of ~1 V across porous carbon electrodes removes salt from feedwaters via electrosorption. In flow-through electrode (FTE) CDI cell architecture, feedwater is pumped through macropores or laser perforated channels in porous electrodes, enabling highly compact cells with parallel flow and electric field, as well as rapid salt removal. We here present a one-dimensional model describing water desalination by FTE CDI, and a comparison to data from a custom-built experimental cell. The model employs simple cell boundary conditions derived via scaling arguments. We show good model-to-data fits with reasonable values for fitting parameters such as the Stern layer capacitance, micropore volume, and attraction energy. Thus, we demonstrate that from an engineering modeling perspective, an FTE CDI cell may be described with simpler one-dimensional models, unlike more typical flow-between electrodes architecture where 2D models are required.

## Introduction

Capacitive deionization (CDI) is a rapidly growing research field, with primary applications in brackish water desalination and wastewater purification.[1] A CDI cell typically consists of two carbon-based porous electrodes that are electronically isolated by a separator, and feedwater is pumped through the cell. Applying a voltage across the electrodes causes charged ions in the feed to migrate to oppositely charged electrodes and to be electrostatically contained in electric double layers (EDLs) within micropores.[2,3] This process constitutes the charge half-cycle, and is also the desalination stage. Once the electrodes are fully charged, they can be discharged by short circuiting the electrodes, allowing the stored ions to be released into the flow and resulting in a waste brine stream. A number of CDI cell architectures have been developed,[4–7] but the earliest and most common architecture is composed of two electrodes separated by a separator channel, through which the feed water is pumped. This architecture is often referred to as flow-by or flow-between electrodes (FB).[1]

An alternative CDI cell architecture is the flow-through electrodes (FTE) architecture, where the feedwater flows directly through electrode macropores rather than between the electrodes (see Figure 1A).[8–10] One main advantage of FTE relative to FB is that the electric field and flow directions are parallel, allowing for facile optimization of ionic and flow resistances[8]. Further, since the separator is no longer the main flow channel in an FTE cell, the separator thickness may be minimized (provided the electronic isolation remains adequate), resulting in improved desalination rates and more compact cells.[8,11] However, it has been reported that anode corrosion occurs at a faster rate in FTE CDI systems relative to FB systems, though nitrogen sparging to reduce dissolved oxygen content in the feedwater has been shown to increase FTE cell stability to a level comparable to FB cells.[12] Further, surface charge modification has been shown to reduce anode corrosion and improve charge efficiency in FTE systems.[9] Another potential drawback of FTE is that such cells can require greater feed pressures than FB cells in order to flow through the electrodes' macropores with the desired throughput.[8] However, recent work by Guyes et al. has demonstrated that laser perforating electrodes with roughly 200 $\mu$m diameter flow channels enabled orders of magnitude improvement in electrode hydraulic permeability without affecting the electrodes salt adsorption capacity or gravimetric capacitance.[13]

Several engineering models for water desalination by CDI have been proposed which couple macroscopic porous electrode theory to an EDL structure model.[14–18] The models developed to date are generally applied to flow-between CDI cells, where flow and electric field are perpendicular, necessitating a 2D model approach.[17,18] Hemmatifar et al. demonstrated the first fully 2D model for flow-between CDI cells, which employed a Donnan EDL model.[18] A widely-applied model utilizes a modified Donnan theory to describe the EDL in micropores of CDI electrodes, and which demonstrates good fits to data over a wide range of experimental conditions and electrode materials.[19,20] While FTE CDI is a promising CDI cell architecture, to our knowledge there has not been a comparison between FTE CDI data to an appropriate model. We here develop a 1D model and simplified boundary conditions for FTE CDI cells, employing a modified Donnan EDL model. We further present the fitting of our model to FTE CDI data from a custom-built cell.

# Theory

To develop a 1D FTE CDI model, we start with the volume-averaged, 1D, superficial molar flux of an ion, $J_i$, given by the extended Nernst-Planck equation,

$$J_i = c_{mA,i} \cdot v - D_{mA,i} \cdot \left( \frac{\partial c_{mA,i}}{\partial x} + z_i c_{mA,i} \frac{\partial \phi_{mA}}{\partial x} \right) \tag{1}$$

where $c_{mA,i}$ is the ion concentration in the macropores of the electrode (defined as a concentration per unit macropore volume), $v$ is the superficial fluid velocity of the electrolyte phase, $D_{mA,i}$ is an effective ion diffusion coefficient, $z_i$ is the ion valence, $\phi$ is the dimensionless macropore electric potential (which can be multiplied by the thermal voltage $V_T=RT/F$ to arrive at a dimensional voltage), and $x$ is a spatial coordinate along the flow and electric field direction in our model FTE CDI cell (see Fig. 1). The effective ion diffusion coefficient in the electrodes, $D_{mA,i} = p_{mA} D_{\infty,i}/\tau_{mA}$, where $D_{\infty,i}$ is the ion's molecular diffusivity, includes a correction for macropore porosity, $p_{mA}$, and tortuosity, $\tau_{mA}$. For simplicity, we assume a binary electrolyte with univalent ions and equal cation and anion diffusivities, whereas future works will investigate the effect of more complex electrolyte solutions.

A conservation of species applied to anion or cation yields

$$\frac{\partial c_{eff,i}}{\partial t} = -\frac{\partial J_i}{\partial x} \quad , \quad c_{eff,i} = p_{mA} c_{mA,i} + p_{mi} c_{mi,i} \tag{2}$$

where $p_{mi}$ is the porosity of the electrode's micropores. We combine Eqs. (1) and (2) to arrive at salt and charge balance equations, given by

$$\frac{\partial c_{eff}}{\partial t} = -v \cdot \frac{\partial c_{mA}}{\partial x} + D_{mA} \cdot \frac{\partial^2 c_{mA}}{\partial x^2} \quad , \quad c_{eff} = p_{mA} c_{mA} + \tfrac{1}{2} p_{mi} c_{mi,ions}$$

$$p_{mi} \frac{\partial \sigma_{ionic}}{\partial t} = 2 \cdot D_{mA} \cdot \frac{\partial}{\partial x} \left( c_{mA} \frac{\partial \phi_{mA}}{\partial x} \right) \tag{3}$$

where $c_{mA}$ is the macropore salt concentration ($= c_{mA,+} = c_{mA,-}$ by electroneutrality), $c_{mi,ions}$ is the total ion concentration in the micropores ($= c_{mi,+} + c_{mi,-}$), and $\sigma_{ionic}$ is the ionic micropore charge ($= c_{mi,+} - c_{mi,-}$).

Micropores in porous CDI electrodes are responsible for salt electrosorption and present a highly confined geometry. One method for modeling the EDL structure within such confined geometry is a Donnan or modified Donnan approach.[21] In the Donnan approach, the potential in the micropore volume is assumed to be constant, independent of the distance to the carbon wall. Furthermore, assuming that ion transport between micropores and macropores (those at the same $x$-position) is rapid and so transport across the electrode thickness is rate limiting, Boltzmann's law relates ion concentrations in micro- and macropore volumes,

$$c_{mi,i} = c_{mA,i} \cdot \exp(-z_i \cdot \Delta \phi_D + \mu_{att}) \tag{4}$$

where $\Delta \phi_D$ is the (dimensionless) Donnan potential, defined as the potential within the volume of micropores relative to that in adjacent macropores. An empirical ion attraction term $\mu_{att}$ is used which aids in fitting of the theory to data ($\mu_{att}$ is assumed to be the same for both ions), which is an inverse function of total micropore ions concentration, $\mu_{att} = E/c_{ions,mi}$, with $E$ a constant micropore attraction energy.[22] This approach has the advantage of relative mathematical simplicity and a good fit of data to theory.[20,22] More recent theories model the EDL structure without the use of a term $\mu_{att}$, instead including charged surface groups in the micropores, termed an amphoteric Donnan model.[23] We here do not employ the latter approach, which can be explored in a future work that also experimentally determines the surface groups and surface group charge density in FTE micropores.

For the modified Donnan EDL model, mobile ionic charge in the micropores, $\sigma_{ionic}$, is equal in magnitude to the electronic charge, $\sigma_{elec}$, which resides in the carbon matrix surrounding the micropore, $\sigma_{ionic} + \sigma_{elec} = 0$. When anode and cathode have the same size and microporosity, the thickness-averaged electronic charge in one electrode is equal in magnitude to the average electronic charge in the other electrode: $\langle\sigma_{elec,A}\rangle + \langle\sigma_{elec,C}\rangle = 0$. In this case, we can relate the ionic current density in the separator layer, $J_{ch}$, to the averaged electrode charge as

$$p_{mi} \frac{\partial \langle\sigma_{elec,j}\rangle}{\partial t} = \pm \frac{J_{ch}}{L_{elec}} \tag{5}$$

where the sign, +/-, depends on the electrode. The EDL model is completed with the following equations, solved together with the PDEs (Eqs. 3) at each $x$-position,

$$c_{ions,mi}^2 = \sigma_{ionic}^2 + \left(2c_{mA}e^{\mu_{att}}\right)^2 \tag{6}$$

$$\phi_1 - \phi_{mA} = \Delta\phi_D + \Delta\phi_S = -\sinh^{-1}\left(\sigma_{ionic}/\left(2c_{mA}e^{\mu_{att}}\right)\right) - \sigma_{ionic} \cdot F/(C_S V_T)$$

where in the above equations, $C_S$ is the volumetric Stern layer capacitance, and $\phi_1$ is the solid phase (carbon) potential. The cell voltage is $V_{cell} = V_T \cdot (\phi_{1,A} - \phi_{1,C})$, where A and C refer to anode and cathode.

For the spacer, we use Eqs. (3) with $p_{mi} = 0$ and $p_{mA}$ replaced by $p_{sp}$. Note that the effective diffusion coefficient is different in the electrode and spacer due pore structure differences, see Table 1. At the electrode-spacer interface, $x = l_e$ and $x = l_e + l_{sp}$, where $l_e$ is the electrode thickness and $l_{sp}$ is the spacer thickness (see Fig 1a), the continuity of salt flux results in

$$D_{mA} \frac{\partial c_{mA}}{\partial x}\bigg|_{x=l_e, l_e+l_{sp}} = D_{sp} \frac{\partial c_{sp}}{\partial x}\bigg|_{x=l_e, l_e+l_{sp}}. \tag{7}$$

Because of continuity of the current, the current density across the spacer, $J_{ch}$, is given by

$$J_{ch} = -2D_{sp}c_{sp} \frac{\partial \phi_{sp}}{\partial x}. \tag{8}$$

Eq. (8) can be integrated to yield

$$J_{ch} \cdot \int_{l_e}^{l_e+l_{sp}} c_{sp}^{-1} \, dx = -2D_{sp} \cdot \left[\phi_{sp}(x = l_e + l_{sp}) - \phi_{sp}(x = l_e)\right]. \tag{9}$$

To derive boundary conditions for the upstream end of our FTE CDI cell ($x = 0$), we begin with a balance of salt applied to a long upstream reservoir,

$$\frac{\partial}{\partial t}\left[\int_{res} c_{res}\, dx\right] = v\left[c_{feed} - c_{mA}(x=0)\right] + D_{mA}\left.\frac{\partial c_{mA}}{\partial x}\right|_{x=0}. \tag{10}$$

Since the reservoir is long, the concentration at the upstream end of the reservoir is unperturbed by the desalination process and remains fixed at the feed concentration, $c_{feed}$. In Eq. 10, the integral in the left-hand side is over the length of the upstream reservoir, and $c_{res}$ is the local concentration in the upstream reservoir. Eq. 10 takes into account the advection of salt into and out of the reservoir, the diffusion of salt into the upstream electrode of the FTE CDI cell, and local changes in concentration in the reservoir as a result of salt diffusion. A concentration boundary layer forms in the reservoir at the upstream reservoir/electrode interface due to the diffusion of salt into the salt-depleted electrode pore space. Taking the limit of high Peclet number, $Pe = vl_e/D_{mA} \gg 1$, the concentration boundary layer thickness, $\delta$, becomes much smaller than the geometric length scale so that $\varepsilon \equiv \delta/l_e \ll 1$. If we now restrict the reservoir domain to only the boundary layer, we can scale Eq. 10 using $c^* \equiv c/c_{feed}$, $t^* \equiv t D_{mA}/l_e^2$, $x^* \equiv x/\delta$, to obtain

$$Pe^{-1}\frac{\partial}{\partial t^*}\left[\int_{res} c_{res}^*\, dx^*\right] = \frac{1}{\varepsilon}\left[1 - c_{mA}^*(x=0)\right] + \frac{1}{\varepsilon^2 Pe}\left.\frac{\partial c_{mA}^*}{\partial x^*}\right|_{x=0}. \tag{11}$$

The time scaling used in Eq. (11) is the characteristic timescale for desalination by a CDI cell,[24] as this desalination is what drives salt removal from the upstream reservoir. For $Pe \gg 1$ and so $\varepsilon \ll 1$ (for our experimental cell, $Pe \sim 25$), we can neglect the left-hand side term in Eq. (11). Thus, we obtain a simple boundary condition for concentration which we employ at the upstream end of the model cell, $x = 0$,

$$0 = v\left[c_{feed} - c_{mA}(x=0)\right] + D_{mA}\left.\frac{\partial c_{mA}}{\partial x}\right|_{x=0}. \tag{12}$$

On the downstream end of the cell ($x = 2l_e + l_{sp}$), salt transport between the electrode and downstream reservoir is due to advection and diffusion. Unlike at the upstream end, a thin concentration boundary layer is not expected to form in the downstream reservoir. As a result, for conditions of high Pe, diffusive flux of salt at the interface is much smaller than advective flux. Thus, we here neglect the diffusive flux, and apply a boundary condition at the downstream end of the cell of $\partial c_{mA}/\partial x|_{x=2l_e+l_{sp}} = 0$. The boundary conditions applied at the upstream and downstream ends of the cell for potential are $\partial \phi_{mA}/\partial x|_{x=0, 2l_e+l_{sp}} = 0$, as no ionic current leaves or enters the cell and we assumed equal ion diffusivities.

Finally, we also include in our model a mixing tank and a plug flow reactor downstream of the cell in order to capture the effect of a significant excess volume before the conductivity sensor (~1 mL for our experimental system, see Materials and methods section). In this volume, mixing and dispersion can act to reduce concentration gradients, affecting cell effluent concentration measurements. The model for the stirred tank after the cell is given by

$$t_{mix} \frac{\partial c_{cs}}{\partial t} = c(x = 2l_e + l_{sp}) - c_{cs} \tag{13}$$

where $t_{mix}$ is the average residence time in the mixing tank (given by its volume divided by volumetric flow rate), and where $c_{cs}$ is the concentration as sensed by the conductivity sensor. The plug flow reactor simply institutes a time-delay, $t_{plug}$, of the effluent conductivity profile.

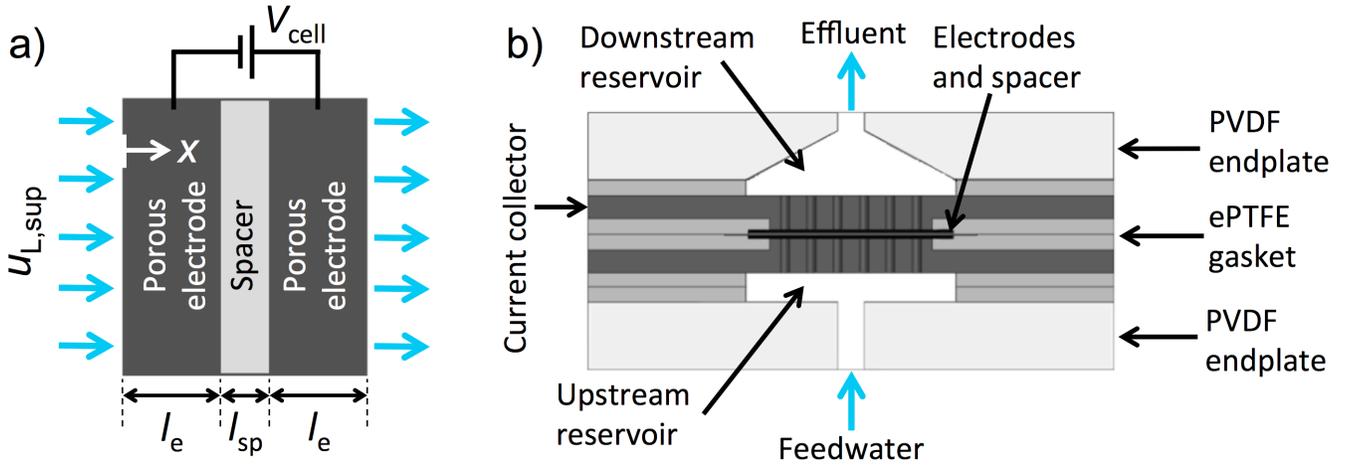

**Figure 1**: a) Schematic of the 1D model domain, which includes both electrodes and the spacer. b) Schematic of the experimental FTE CDI cell used in this work, with cell dimensions provided in the Materials and Methods section.

Table 1. List of model parameters and their values.

| Model parameter | Value |
|---|---|
| $\rho_{elec}$: electrode mass density | 0.25 g/mL |
| $v_{mi}$: Specific volume of micropores | 0.55 mL/g (fitting parameter) |
| $p_{mi}$: microporosity | $\rho_{elec} \cdot v_{mi} = 0.1375$ |
| $p_{sk}$: "carbon skeleton" | $\rho_{elec} / \rho_{sk} = 0.1316$ ($\rho_{sk} \sim 1.9$ g/mL) |
| $p_{mA}$: mAcroporosity | $1-p_{mi}-p_{sk}=0.7309$ |
| $E$: ion attraction energy | 700 "kT"·mol/m$^3$ (fitting parameter) |
| $C_S$: Stern capacity | 145 F/mL (fitting parameter) |
| $l_e$: electrode thickness | 500 μm |
| $l_{sp}$, $p_{sp}$: spacer thickness and porosity | 260 μm, 0.85 |
| $u_{L,sup}$: superficial velocity | 66.4 μm/s |
| $t_{pc-mv}$: mixing vessel retention time | 60 s |
| $t_{plug}$: plug flow reactor time | 15 s |
| $D_\infty$: ion diffusivity | $D_\infty=(D_{Na}+D_{Cl})/2=1.68\cdot10^{-9}$ m$^2$/s |
| $D$: effective diffusion coefficients | $D_{mA}=D_\infty \cdot p_{mA}/\tau_{mA}$, $D_{sp}=D_\infty \cdot p_{sp}/\tau_{sp}$ |
| $\tau$: Tortuosity | $\tau_{mA/sp}=1/p_{mA/sp}^{1/2}$ (Bruggeman equation) |

# Materials and methods

The FTE CDI cell (Figure 1b) consists of a pair of commercial porous activated carbon woven-fiber electrodes (ACC-507-15, Kynol Europa GmbH, ~500 μm thickness each, 1.75 × 1.75 cm$^2$). A porous separator (GE Life Sciences, Whatman GF/A borosilicate glass filter paper, 260 μm thickness, 2.4 × 2.4 cm$^2$) electronically isolates the electrodes. ePTFE gaskets (W.L. Gore & Associates, Gore-Tex NSG16X-GP, 1.4 mm uncompressed thickness, 5 × 5 cm$^2$) are used to seal the cell, while a laser-cut square in the gasket (1.55 × 1.55 cm$^2$) permits feedwater to pass through the electrodes and separator. The electrodes fit tightly into grooves laser-cut on the aperture perimeter, preventing feedwater from leaking around the electrode edges. The upstream negative current collector is milled from impervious graphite (FC-GR, Graphitestore.com, Inc., Buffalo Grove, IL, USA), and the downstream positive collector is milled from isomolded graphite (GM-10, Graphitestore.com). The current collectors both contain an array of cylindrical channels (6 × 6 grid, 1.5 mm diameter, 3 mm length) that allow water to pass through, and tabs to enable electrical contact with the voltage source. Water reservoirs are created upstream and downstream of the collectors via two ePTFE gaskets on the upstream side and a single ePTFE gasket and pyramidal contraction (the latter is ground by hand with a Dremel cutting tool in the endplate on the downstream side). The downstream reservoir components have a combined volume of ~0.5 mL. The cell terminates on either side with endplates milled from PVDF (5 × 5 cm$^2$) that each include one fluid flow line and one vent that allows air removal from the cell. Effluent enters the outlet line with a tube volume totaling 0.5 mL between the endplate and downstream conductivity sensor. The cell is sealed with 8 bolts (M4 × 30 mm) tightened to 35 N·cm, which are electrically isolated from the graphite current collectors via shrink tubing.

For desalination experiments, feedwater (NaCl 5 mM or 20 mM concentration) was drawn at 1 mL/min from a glass bulk reservoir by the peristaltic pump and fed into the CDI cell via semi-rigid polyethylene tubing. The cell effluent was then fed into a conductivity cell (5-ring, Metrohm, Inc., Switzerland) with a custom-milled insert to reduce the internal volume, and then returned to the bulk reservoir. The bulk reservoir volume (~2 L) was significantly larger than the volume of the rest of the setup (~20 mL) in order to maintain constant concentration in the reservoir throughout experiments. During the charging half-cycle of the desalination experiments, a constant voltage of between 0.2 and 1.2 V was applied to the cell by a voltage source (SourceMeter 2400, Keithley Instruments, Inc., Solon, OH, USA), while during the discharging half-cycle a cell voltage of 0 V was maintained.

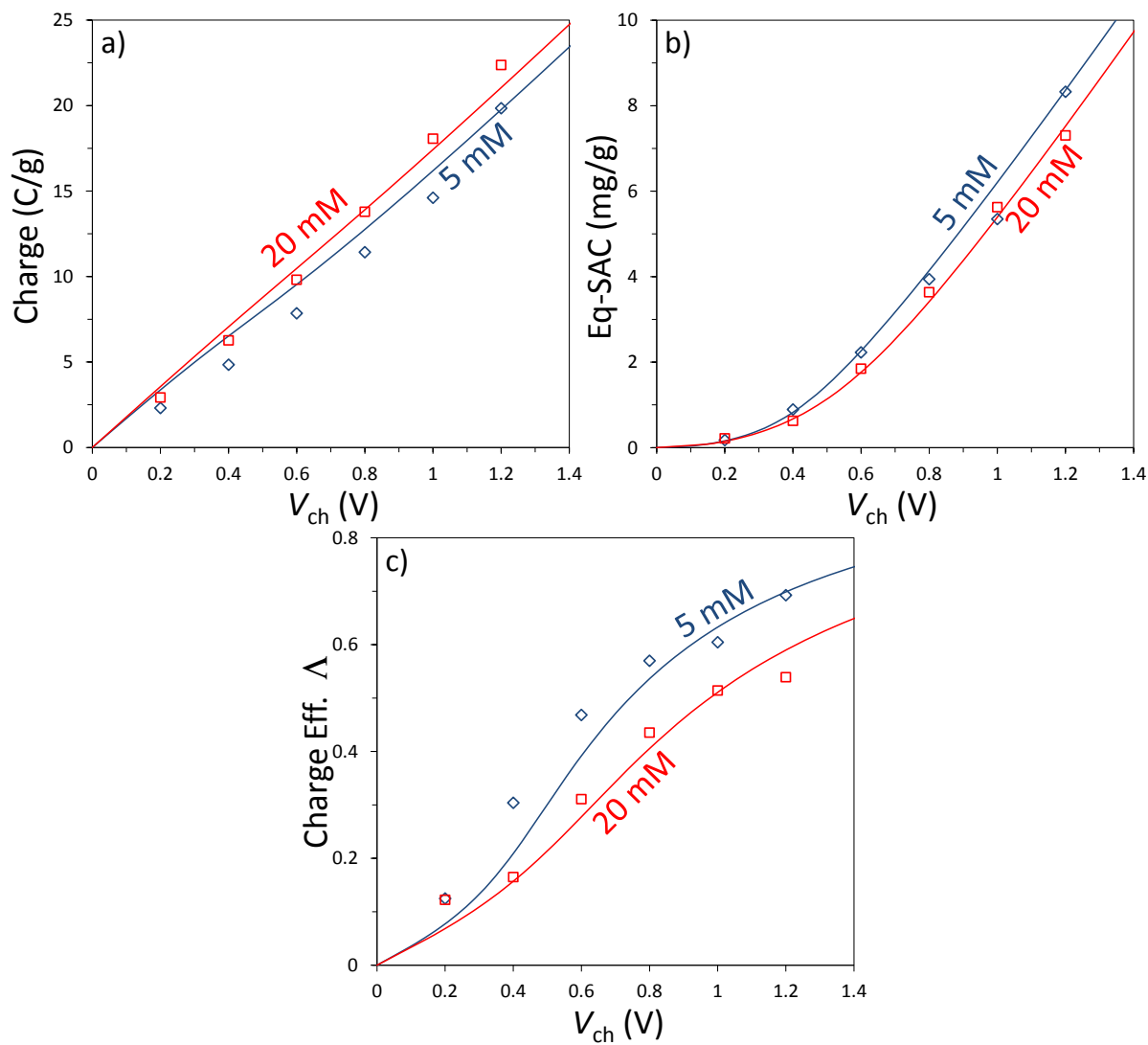

**Figure 2**: Results of model-to-data fitting for several measured cell parameters at equilibrium including a) charge stored, b) equilibrium salt adsorption capacity (eq-SAC), and c) charge efficiency, as function of charging voltage $V_{ch}$.

# Results and Discussion

The first step towards comparison of data to the model is to fit equilibrium data of charge stored and salt adsorption capacity (eq-SAC) to the equilibrium results of the model. Data was taken with our custom-build FTE CDI cell operated in constant voltage mode, where the charge half-cycle was continued until equilibrium. Equilibrium occurred when the effluent concentration returned to the feed concentration and the current decreased to reach a steady value (leakage current). Charge stored was obtained by integrating the cell's current response during the discharge half-cycle. Eq-SAC was obtained by integrating the difference between feed and effluent concentration during the charge half-cycle, multiplied by the feed flow rate.

The results of the fitting are shown in Figure 2a and b, and as can be seen, model-to-data fitting gave good agreement for fitting parameters of $v_{mi} = 0.55$ g/mL, $E = 700$ kT·mol/m$^3$ and $C_S = 145$ F/mL. These values are similar to those obtained from equilibrium model-to-data fitting for other CDI electrode materials, though the value for $E$ is at least twice higher than previously reported.[22,25] The latter may be due to slight variations in the micropore size of the electrode material used here compared to those used previously, since the parameter $E$ is expected to scale as $\lambda_p^{-4}$, where $\lambda_p$ is micropore size.[22] Figure 2c also shows a model-to-data comparison for the equilibrium value of the parameter charge efficiency, which is defined as the moles of salt stored in the electrodes to the moles of electrons stored. Charge efficiency of our experimental cell varied from ~0.1 at a cell voltage of 0.2 V to roughly 0.7 at 1.2 V for the case of 5 mM feed concentration, similar to the trends predicted by the model.

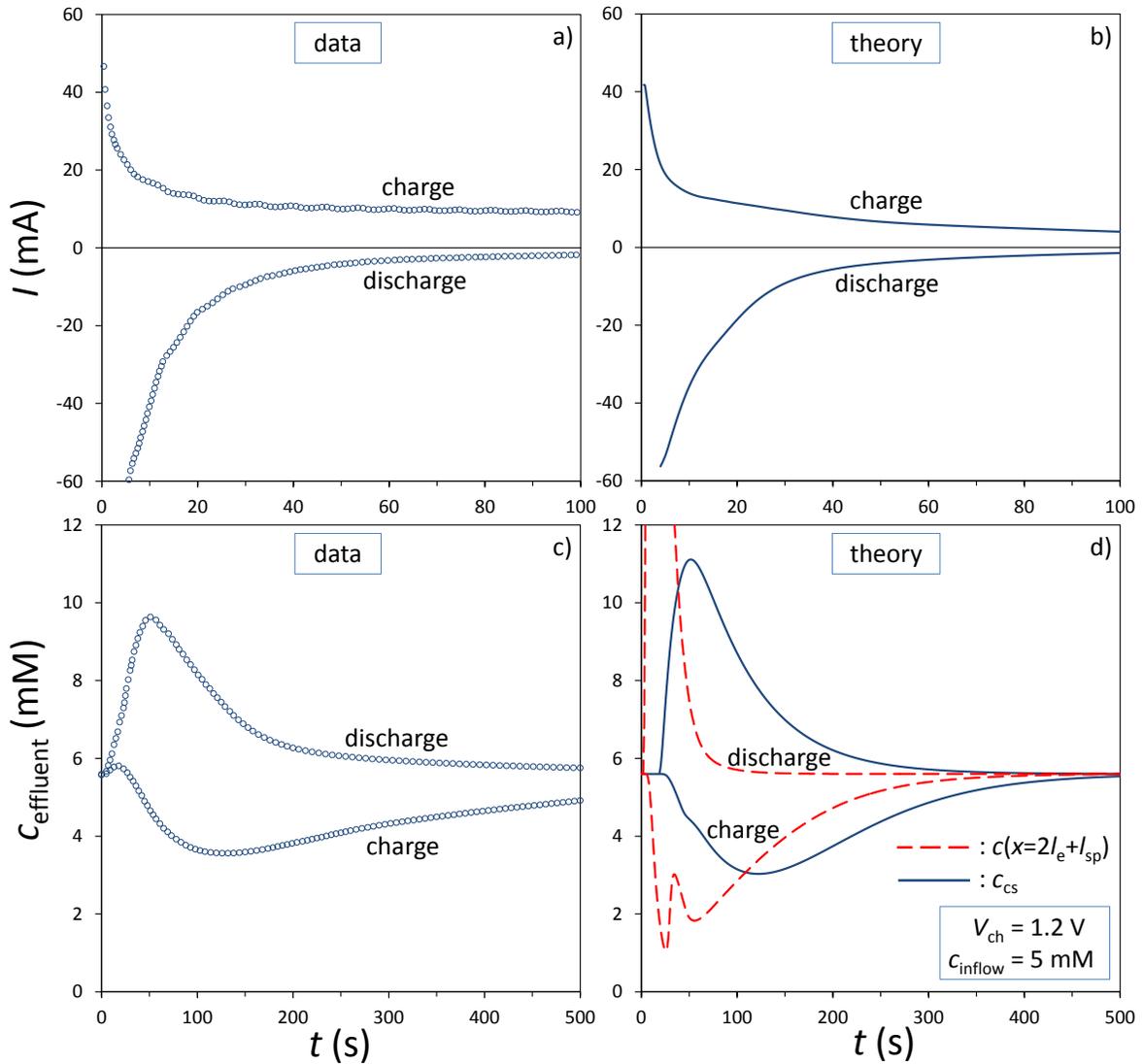

**Figure 3**: Comparison of model results to data for the charge half-cycle, $c_{inflow}$=5 mM NaCl, and $V_{ch}$=1.2 V. a) and b) represent the comparison for current response of the CDI cell, while c) and d) are for the cell effluent concentration. The red dashed lines in d) represent the effluent concentration predicted by the model at the exit of the downstream electrode of the CDI cell.

After determining the fitting parameters from equilibrium model-to-data fitting, we could compare dynamic model and data results, which is done in Figure 3 for the case of 5 mM NaCl feedstream and Figure 4 for the case of 20 mM NaCl feedstream. In Figure 3 and 4, we can see good qualitative agreement between model results and data when including the effect of the downstream volume in our cell (Eq. 13). Good agreements are obtained for both the cell's current response ($I$, Fig 3a and b and 4a and b) and effluent concentration ($c_{\text{effluent}}$, Fig 3c and d and 4c and d), and for both the charge and the discharge half-cycles.

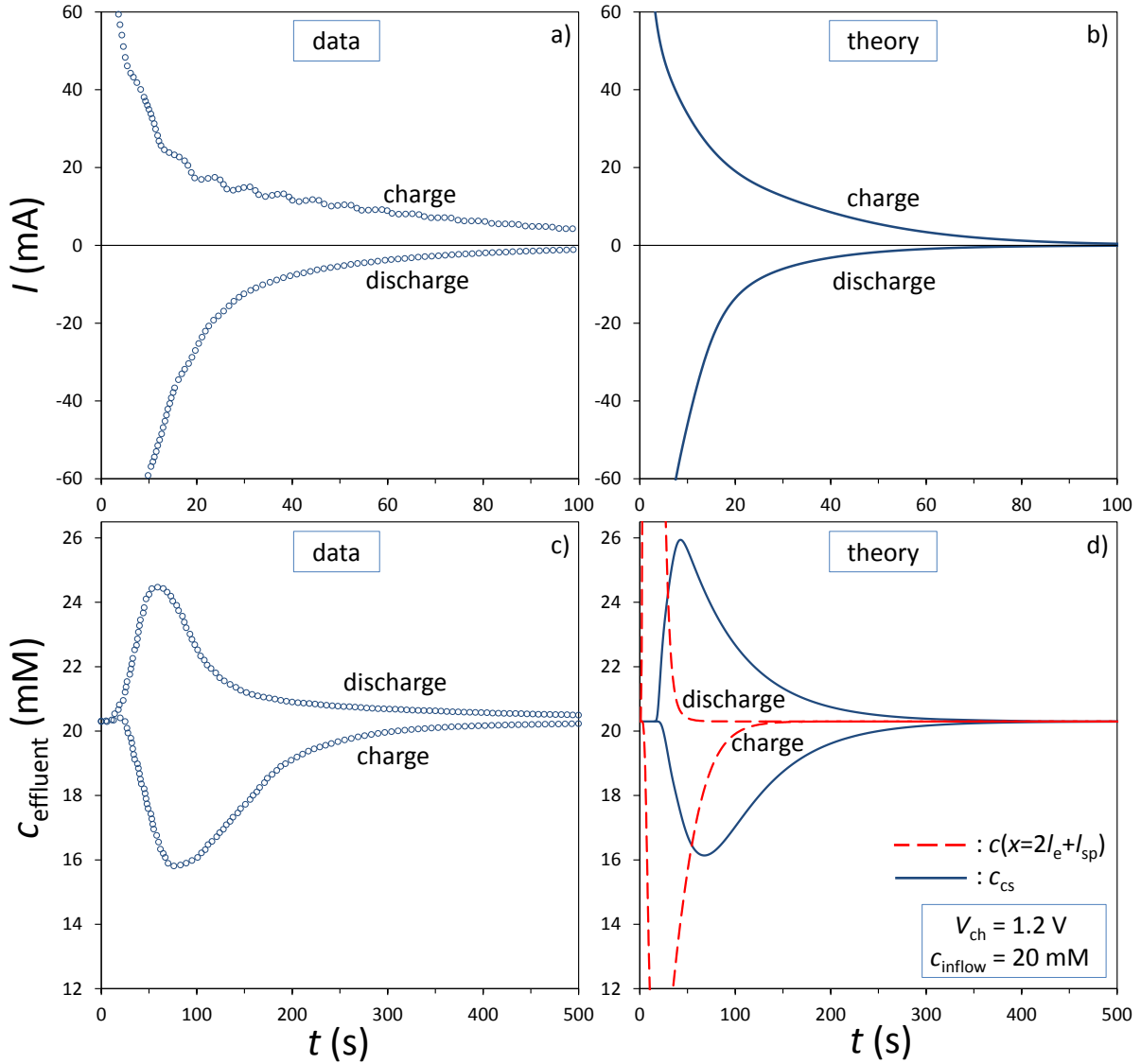

**Figure 4**: Comparison of model results to data for the charge half-cycle, $c_{\text{inflow}} = 20$ mM NaCl, and $V_{\text{ch}} = 1.2$ V. a) and b) represent the comparison for current response of the CDI cell, while c) and d) are for the cell effluent concentration. The red dashed lines in d) represent the effluent concentration predicted by the model at the exit of the downstream electrode of the CDI cell.

Also shown in Figure 3d and 4d is the predicted effluent concentration when not including the effect of the downstream volume in the model (red lines), which shows sharper features and poor comparison to the experimental data (Fig 3c and 4c). The latter demonstrates the importance of accounting for the electrolyte volume and diffusion effects downstream of the cell in predicting the effluent concentration at the location of the conductivity sensor. Further, the results of Figs. 3 and 4 show that a 1D model can be used to predict experimental data from FTE CDI cells. In FTE CDI cells, flow and electric field are parallel, unlike flow-between cells, which necessarily require a 2D model as flow and electric field are perpendicular.

# Conclusions

In conclusion, we present a model for flow-through electrode capacitive deionization (FTE CDI) based on modified Donnan EDL structure model and porous electrode transport theory. Although possessing similarities to models for flow-between CDI cells, our model was unique in that a simple one-dimensional approach was able to capture experimental results, and simple boundary conditions derived via scaling arguments were developed. Model-to-data comparisons showed good qualitative agreement when including the effect of a significant volume downstream of the cell. Future work can model other features important to desalination by FTE CDI, such as pH variations in the cell effluent and the effects of charged surface groups in porous electrodes.

# Acknowledgements

We would like to acknowledge funding from the Israeli Ministry of National Infrastructures, Energy and Water Resources.